\documentclass[aps,prb,twocolumn,groupedaddress,showpacs]{revtex4}
\usepackage{graphicx}
\usepackage{amssymb}
\usepackage{bm}
\usepackage{subfigure}
\usepackage{graphpap}
\begin{document}
%%%

%-----------------------------------------------------------------------
\title{ {\em Ab-initio} transport properties of nanostructures from
maximally-localized Wannier functions }
%-----------------------------------------------------------------------
\author{Arrigo Calzolari, $^1$ 
Nicola Marzari,$^2$ Ivo Souza,$^3$ and
Marco Buongiorno Nardelli$^4$ }
%-----------------------------------------------------------------------

\affiliation{ $^1$ INFM-S$^3$ - National Research Center on
nanoStructures and Biosystems at Surfaces, and Dipartimento di Fisica
Universit\`a di Modena e Reggio Emilia, I-41100 Modena, Italy\\ 
$^2$ Department of Materials Science and
Engineering, Massachusetts Institute of Technology, Cambridge,
Massachusetts 02139-4307\\ 
$^3$ Department of Physics and Astronomy, Rutgers University, Piscataway,
New Jersey 08854-8019\\ 
$^4$ Department of Physics, North Carolina State University, Raleigh, NC 27695 and CCS-CSM, Oak Ridge National Laboratory, Oak Ridge, TN 37831}
%-----------------------------------------------------------------------
%%%%%%%
%%%%%%%
\date{\today}

\begin{abstract}
We present a comprehensive first-principles study of the ballistic
transport properties of low dimensional nanostructures such as linear
chains of atoms (Al, C) and carbon nanotubes in presence of defects.
A novel approach is introduced where quantum conductance is 
computed from the combination
of accurate plane-wave electronic structure
calculations, the evaluation of the corresponding maximally-localized
Wannier functions, and the calculation of transport properties by a
real-space Green's function method based on the Landauer
formalism. This approach is computationally very efficient, can
be straightforwardly implemented as a post-processing step in a
standard electronic-structure calculation, and allows to directly 
link the electronic transport properties of a device to the 
nature of the chemical bonds, providing insight onto
the mechanisms that govern electron flow at the nanoscale.
 
\end{abstract}

\maketitle

%%%%%%%%%%%%%%%%%%%%%%%%%%%%%%%%%%%%%%%%%%%%%%%%%%%%%%%%%%%%%%%%%%%%%%
%%%                  Introduction                                 %%%%
%%%%%%%%%%%%%%%%%%%%%%%%%%%%%%%%%%%%%%%%%%%%%%%%%%%%%%%%%%%%%%%%%%%%%%

\section{Introduction}

The field of nanotechnology has undergone a remarkable growth in the
last few years.  This development has been fueled by the expectation
that unusual properties of matter,~\cite{nygard,park,liang} which
become evident as the dimensions of the structural components of a
device shrink under $\sim$10-100 nm, may be exploited.  Indeed,
the challenges for future
developments involve continuous shrinking of the physical
dimensions of devices and attainment of higher speeds.  The
drive to produce smaller devices has led the current research towards
new forms of electronics, in which nanoscale objects and molecular
devices replace the transistors of today's silicon
technology.~\cite{aviram,joachim,williams,gudiksen} Experiments have
been performed to directly measure charge transport properties of
hybrid metal-(bio)molecular systems
\cite{aviram,joachim,reed,collier,metzger,braun,reichert,smit,lahann}
and carbon-based aggregates such as fullerenes and
nanotubes.~\cite{williams,nygard,tans,white,ouyang,collins} Particular
attention has been devoted to atomic-scale devices,
since they represent the limit towards one-dimensional
electronics, and thus the transport
properties of wire-like chains of atoms (especially Au, Al and C),
connected with metal electrodes have been widely
investigated.~\cite{agrait, ohnishi,yanson,nilius}  Despite their
simple structures, atomic-sized chains display peculiar quantum
properties due to their low dimensionality; in particular, the electronic
properties are strongly affected by the nature of single chemical
bonds and coordination numbers.

The ongoing rapid advances in the measurements of electrical
conductance in individual molecular- and atomic-sized devices require
commensurate advances in the theoretical understanding of the detailed
microscopic mechanisms that control charge mobility.
Modeling of single nano-elements and coupled arrays of
nano-devices is needed to provide interpretation and feedback to
experimental measurements, to predict device characteristics, and to
provide a basis for the functional progress of these new devices.

In general, the electron transport properties of nanostructures can be
simply described in the Landauer formalism.~\cite{landauer,datta} The
Landauer relation connects the quantum conductance ${\mathcal G}$ with
the transmission function ${\mathcal T}(E)$: its evaluation requires
the knowledge of the electronic structure of the system under
consideration and the inclusion of scattering at
contacts.  This approach relies on the evaluation of lattice 
Green's functions of the system. Several approaches have been
developed to calculate the quantum conductance in nanostructures,
based on semi-empirical (tight-binding, H\"uckel)
models~\cite{marco,todorov,kristic,rego}; more recently, a
variety of first-principles formulations have
appeared.~\cite{lang1,choi,landman,landman1,yoon,guo,yaliraki,stokbro,palacios,datta1}
{\em Ab-initio} approaches have also been extensively used to
characterize the electrical properties of nano- and
bio-materials,~\cite{arrigo,parrinello1} and to study
the effects of microscopic structural relaxation
and of electrode/conductor junctions.~\cite{rosa}

In this paper, we present an original approach to the calculation of
coherent transport properties of nanostructures from first principles. Our
methodology combines an accurate description of the
electronic ground state provided by well-developed first-principles
calculations based on plane-wave (PW) representations, with the
Landauer approach to describe transport properties of extended
systems.~\cite{marco1,marco2} The essential connection is
provided by the use of the maximally-localized Wannier function
representation~\cite{nicola} that allows  to introduce naturally the 
ground-state electronic structure 
into the lattice Green's function approach that will be our
basis for the evaluation of the Landauer quantum conductance.

The paper is organized as follows.  In Section~\ref{method} we
describe the main features of the method.  In Section~\ref{bulk-like}
we will study the {\em bulk} electronic and conduction properties of
linear chains of Al and C atoms.  Section~\ref{lead-like} deals with
the conduction properties of a carbon nanotube with a substitutional
defect. This example is used to elaborate on the formulation of the
two-terminal conductance problem in our approach.  The paper ends with
Section ~\ref{conclusions} where our conclusions are presented.

%
%
%%%%%%%%%%%%%%%%%%%%%%%%%%%%%%%%%%%%%%%%%%%%%%%%%%%%%%%%%%%%%%%%%%%%%%
%%%                        Method                                 %%%%
%%%%%%%%%%%%%%%%%%%%%%%%%%%%%%%%%%%%%%%%%%%%%%%%%%%%%%%%%%%%%%%%%%%%%%
\section{Method}\label{method}

%%%%%%%%%%%%%%%%%%%%%%%%%%%%%%%%%%%%%%%%%%%%%%%%%%%%%%%%%%%%%%%%%%%%%%
%%%             Electronic Transport                               %%%
%%%%%%%%%%%%%%%%%%%%%%%%%%%%%%%%%%%%%%%%%%%%%%%%%%%%%%%%%%%%%%%%%%%%%%
\subsection{Electronic transport in extended systems}
\label{tbmethod}
Calculations of the quantum conductance are based on a recently
developed efficient method for evaluating quantum transport in
extended systems.~\cite{marco,marco1,marco2} This method is applicable
to any Hamiltonian that can be expanded within a localized-orbital
basis and can be used as a general theoretical scheme for the
computation and analysis of the electrical properties of
nanostructures.

\begin{figure}[!t]
\includegraphics[clip,width=0.40\textwidth]{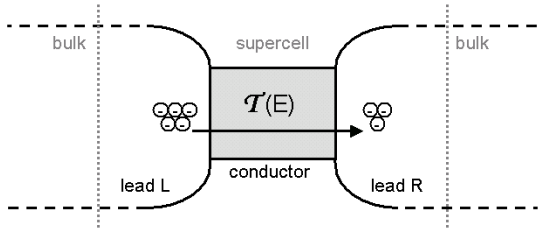}
\caption{Geometry of a typical conductance calculation.  Charge
carriers can be transmitted throughout the contact region (grey) from
the left to the right lead.  Dotted lines separate the device in the
three regions (left lead/conductor/right lead) that enter the
conductance calculation. While the leads can be easily described with
a primitive periodic unit cell,
the conductor region has to be described with a supercell calculation
that includes -- along the direction of electron transmission -- both the
conductor and the lead-conductor contact region (see Sec. IV).}
\label{lcr}
\end{figure}

Let us consider a system composed of a conductor
connected to two semi-infinite leads, as in Fig.~\ref{lcr}.  The
quantum description of the electronic conductance is a complex
non-equilibrium problem. We begin the study of conduction properties
focusing on the {\em coherent electron transport}.  This approach
leaves out non-equilibrium effects due, e.g., to dissipative
scattering or to an external bias.\\
\noindent
Quantum conductance (${\mathcal G}$) is the microscopic quantity 
that characterizes the transport properties of a conductor 
and may be calculated using the Landauer expression:~\cite{landauer}
\begin{equation}
{\mathcal G}(E) = {2 e^2 \over h} {\mathcal T}(E),
\label{eq:landauer}
\end{equation}
where ${\mathcal T}$(E) is the transmission function, i.e.
the probability that an electron injected at one end of the conductor
with energy E will be transmitted to the other end.  In the Green's function
formalism the transmission function ${\mathcal T}$ is expressed
as \cite{datta}
\begin{equation}
{\mathcal T} = {\rm Tr}(\Gamma_L G_C^r \Gamma_R G_C^a),
\label{eq:T}
\end{equation}
where $G_C^{\{r,a\}}$ are the retarded and advanced Green's
functions of the conductor, and $\Gamma_{\{L,R\}}$ 
represent the coupling of the conductor to the leads.

The effect of the semi-infinite leads on the conductor can be
described by means of finite-dimension 
operators known as the {\em self-energies}
$\Sigma_{L,R}$.  The Green's function $G_C$ and the coupling functions
$\Gamma_{\{L,R\}}$ of Eq.~\ref{eq:T} are explicitly obtained from
the self-energies as:
\begin{equation}
G_C = (\epsilon -H_C -\Sigma_L -\Sigma_R)^{-1}, 
\label{gconduct}
\end{equation}
\begin{equation}
\Gamma_{\{L,R\}} = {\rm i}[\Sigma_{\{L,R\}}^r -
\Sigma_{\{L,R\}}^a]. 
\label{eq:gamma}
\end{equation}
where $H_C$ is the Hamiltonian matrix of the conductor, 
calculated with respect to a localized real-space basis set; 
in our approach {\em all these operators will be
represented in the basis set of the maximally-localized
ground-state Wannier functions (WFs) for the system examined}.
This will allow a mapping for the first-principles Hamiltonian onto an
exact band-structure tight-binding model, for which we then follow the
detailed Landauer formulation of Ref. \onlinecite{marco}.

The key assumption is the description of the original system 
as a sequence of {\em principal layers},~\cite{lee}
along which we calculate the transport properties and where
the interaction between adjacent layers is accounted for by 
transfer matrices $T_{L,R}$ and $\overline T_{L,R}$.
Within this approach the self-energies due to the semi-infinite 
leads are \cite{marco}
\begin{equation}
\begin{array}{l}
\Sigma_L=H_{LC}^{\dagger} (\epsilon
-H_{00}^{L}-(H_{01}^{L})^{\dagger}\overline T_L)^{-1}H_{LC},\\ \\
\Sigma_R=H_{CR}(\epsilon
-H_{00}^{R}-H_{01}^{R}T_R)^{-1}H_{CR}^{\dagger},
\end{array}
\label{sigma}
\end{equation}
where $H_{LC}$, $H_{CR}$ are the coupling matrices for the
conductor-lead assembly, and $H_{nm}^{L,R}$ are the matrix elements of
the Hamiltonian for the infinite (bulk-like) leads. For instance,
$H_{00}^{L,R}$ describes the intralayer interactions and
$H_{01}^{L,R}$ the interlayer coupling between two adjacent layers. If the
orbitals are sufficiently localized, the residual coupling with 
layers farther apart will be negligible;  conversely, the localization
properties of the orbitals spanning the desired energy window 
determine the minimal thickness for a principal layer.
The transfer matrices $\overline T_{L,R}$ and $T_{L,R}$ are also computed from the
Hamiltonian matrix elements via an iterative
procedure.~\cite{garcia}

The only required inputs then are the matrix elements of the Hamiltonian $H_{mn}$
expanded in a localized-orbital basis; by choosing the maximally-localized
WFs representation, we provide essentially an exact mapping
of the ground state onto a minimal basis.~\cite{note_metal}
The accuracy of the results directly depends on having principal layers 
that do not couple beyond next-neighbors,  i.e. on having a well-localized 
basis.

%%%%%%%%%%%%%%%%%%%%%%%%%%%%%%%%%%%%%%%%%%%%%%%%%%%%%%%%%%%%%%%%%%%%%%
%%%             band structure calculations                       %%%%
%%%%%%%%%%%%%%%%%%%%%%%%%%%%%%%%%%%%%%%%%%%%%%%%%%%%%%%%%%%%%%%%%%%%%%
\subsection{Ab-initio electronic structure}\label{pw-bandstructure}
The starting point for our procedure is the first-principles calculation of
the electronic structure of the nanostructure, eventually coupled to the
leads. We adopt a standard electronic-structure method
based on self-consistent total energy and force minimization, which
allows to optimize simultaneously the atomic positions and the
corresponding electronic wavefunctions. The electronic structure is
described within Density Functional Theory
(DFT).~\cite{dreizler} The examples presented in the following Section
are obtained in the Local Density Approximations (LDA),~\cite{pederw}
but more sophisticated corrections at the exchange-correlation
functional ({\em e.g.}  generalized gradient approximation) can obviously be
used.  The electron-ion interaction is described via 
norm-conserving pseudopotentials~\cite{tm} in the  form of
Kleinman and Bylander.~\cite{kb}  We use
Periodic Boundary Conditions (PBC) along the three directions of direct
space and the electronic wavefunctions are expanded in a plane-wave basis 
set compatible with the chosen PBC.
While the translational invariance makes the 
plane waves  a very natural choice to describe the 
wavefunctions of a periodic system, the drawback is
that they are truly delocalized.  Brillouin Zone (BZ) summation
are performed with homogeneous Monkhorst-Pack grids~\cite{mp} of {\bf
k}-points in the the first Brillouin Zone.

This approach to electronic-structure calculations is
widely used, provides a faithful description of the
electronic properties of the systems of interest, and in the present
context it has been successfully
applied to the investigation of both solid-state and molecular
assemblies.~\cite{jones,wimmer}
The results of such calculations are, at each {\bf k}-point, the
Kohn-Sham energy eigenvalues and their corresponding eigenvectors (Bloch states),
expanded in plane waves. It is worth noting that the present
methodology to compute quantum transport from first-principles will apply
to any electronic-structure approach, since it can construct and employ
orbitals that are maximally-localized, and that represent a minimal basis set,
under the single assumption that eigenstates are in the Bloch form.
Even if the basis set used is already localized (albeit not minimal) 
the localization procedure will allow to recover exact results with 
smaller systems. 
In particular, our procedure can be
applied in combination with Car-Parrinello molecular dynamics
simulations, opening the way to compute quantum conductance in large-scale
systems, and as a
function of temperature, completely from 
first-principles.\cite{LeeMBNMarzari}

%%%%%%%%%%%%%%%%%%%%%%%%%%%%%%%%%%%%%%%%%%%%%%%%%%%%%%%%%%%%%%%%%%%%%%
%%%             Maximally Localized Wannier Functions              %%%
%%%%%%%%%%%%%%%%%%%%%%%%%%%%%%%%%%%%%%%%%%%%%%%%%%%%%%%%%%%%%%%%%%%%%%

\subsection{Maximally-localized WFs}\label{wannier_theo}

Bloch orbitals cannot be used directly to evaluate electronic
transport with the method outlined in 
Sec. \ref{tbmethod}. As we have pointed out, 
the quantum conductance is computed starting from the knowledge of 
the lattice Green's
function, whose calculation relies on a localized orbital
representation of the electronic states in real space. Bloch orbitals,
that are intrinsically delocalized, have to be transformed into {\em
localized} functions in order to construct the sparse, short-ranged matrix
elements of the Hamiltonian. 
The core of our proposed methodology
is to use maximally-localized WFs for the system considered. These 
are the most natural choice for a set of localized orbitals 
that still span the same Hilbert space of the Hamiltonian 
eigenfunctions, and they allow  to bridge plane-wave
electronic structure and lattice Green's function calculations in a
coherent fashion. 
In the case of an isolated system the maximally-localized WFs
become Boys localized orbitals;~\cite{boys} 
therefore, our procedure is not
tied to an extended-systems formulation, but can equally well 
represent isolated molecules. (In addition, 
the localization procedure is greatly simplified for the 
case of large unit cells, when $\Gamma$-sampling only is used \cite{Silvestrelli}).

A Wannier function $w_{n{\bf R}}({\bf r})$, labeled by the Bravais
lattice vector {\bf R}, is usually defined via a unitary transformation of
the Bloch functions $\psi_{n{\bf k}}({\bf r})$ of the $n$th 
band:
\begin{equation}
w_{n{\bf R}}({\bf r})=\frac{V}{(2\pi)^3}\int_{BZ}\psi_{n{\bf k}}({\bf r}) 
e^{-i{\bf k}\cdot{\bf R}} d^3k,
\label{wf}
\end{equation}
where V is the volume of the unit cell and the integration is performed over
the entire Brillouin Zone.  It is easy to show that the WFs defined as
above form an orthonormal basis set, and that any two of them, for a given 
index $n$ and different ${\bf R}$ and ${\bf R^\prime}$, are just
translational images of each other.
Note that, as the ${\mathcal N}$ WFs
form a (continuous) linear combinations of Bloch functions with 
different energies, they do not represent stationary states, but 
still span exactly the same original Hilbert space.
The {\em ab-initio} eigenstates are well-defined, modulus an arbitrary
${\bf k}$-dependent phase factor; thus, the definition above
does not lead to a unique set of Wannier functions~\cite{kohn},
since the electronic structure problem is invariant for the transformation 
$\psi_{n{\bf k}} \leadsto e^{\phi_n({\bf k})} \psi_{n{\bf k}} $.
Besides this freedom in the choice of 
phases $\phi_n({\bf k})$ for the Bloch functions,
there is a more comprehensive gauge freedom stemming from the fact that the 
many-body wavefunction is actually a Slater determinant: a unitary
transformation between orbitals will not change the manifold, and
will not change the total energy and the charge density of the system.
In all generality, starting with a set of ${\mathcal N}$ Bloch
functions with periodic parts $u_{n{\bf k}}$, we can constructs infinite 
sets of ${\mathcal N}$ WFs displaying different spatial characteristics:
\begin{equation}
w_{n{\bf R}}({\bf r})=\frac{V}{(2\pi)^3}\int_{BZ}
\left[ \sum_m U_{mn}^{({\bf k})}
\psi_{m{\bf k}}({\bf r}) \right]
e^{-i{\bf k}\cdot{\bf R}} d^3k.
\label{u1}
\end{equation}
The unitary matrices $U^{({\bf k})}$ include also the gauge freedom
on phase factors afore mentioned.~\cite{nicola}

For our purposes, we need to transform the Bloch eigenstates
in WFs with the narrowest spatial distribution.  We
construct {\em maximally-localized WFs} using the
algorithm proposed by Marzari and
Vanderbilt.~\cite{nicola}  We define a {\em Spread Operator}
($\Omega$) as the sum of the second
moments of the Wannier functions corresponding to one choice of
translational lattice vector:
\begin{equation}
\Omega=\sum_n [\langle w_{n{\bf 0}} | r^2 | w_{n{\bf 0}} \rangle - 
\langle w_{n{\bf 0}} | {\bf r} | w_{n{\bf 0}} \rangle^2] ,
\label{omega}
\end{equation}
where the sum is over the group of bands which spans the Hilbert space.
The value of the spread $\Omega$ depends on the choice of unitary matrices
$U^{({\bf k})}$; thus  it is possible
to evolve any arbitrary set of $U^{({\bf k})}$ until the minimum condition
\begin{equation}
\frac {\delta \Omega_{\bf k}}{\delta U^{({\bf k})}}=0 
\label{minimo}
\end{equation}
is satisfied. 
At the minimum, we obtain the matrices $(U^{({\bf k})})^{ML}$ 
that transform the first-principles 
$\psi_{n{\bf k}}^{FP}({\bf
r})$ into the {\em maximally-localized WFs}
$w_{n{\bf R}}^{ML}({\bf r})$:
\begin{equation}
\begin{array}{lll}
\psi_{n{\bf k}}^{ML}({\bf r})&=&\sum_{m} (U_{mn}^{({\bf k})})^{ML}
\psi_{m{\bf k}}^{FP}({\bf r}),\\ \\ w_{n{\bf R}}^{ML}({\bf
r})&=&\frac{V}{(2\pi)^2} \int_{BZ} 
\psi_{n{\bf k}}^{ML}({\bf r}) e^{-i{\bf k}\cdot{\bf R}} d{\bf k} .\\
\end{array}
\end{equation}

A useful feature of the method is that the only ingredients needed to
calculate the spread functional
$\Omega$ and to evolve the unitary matrices $U^{({\bf k})}$ are
the overlap matrix $M_{mn}^{({\bf k},{\bf b})}$
between the periodic part of the Bloch states at neighboring {\bf k}-points:
\begin{equation}
M_{mn}^{({\bf k},{\bf b})}=\langle u_{m,{\bf k}} | u_{n,{\bf k+b}}\rangle,
\label{m}
\end{equation}
where {\bf b} is the vector that links neighboring {\bf k}-points
in the discretized BZ integrals.~\cite{note_ref}

It is important to notice that whenever a Born-von Karman discretization 
of the Brillouin Zone is introduced, even the above-mentioned WFs are not truly
localized, but will be periodic in real-space, with a {\em superperiodicity}
determined by the BZ discretization. The truly isolated limit is
recovered only in the case of continuous BZ integrations.
This is easily seen remembering that 
$ \psi_{n{\bf k}}({\bf r})= u_{n{\bf k}}({\bf r}) 
e^{i{\bf k}\cdot{\bf r}} $, and $  u_{n{\bf k}}({\bf r}) $ has the
periodicity of the direct lattice; thus the phase factors 
$ e^{i{\bf k}\cdot{\bf r}}  $ determine the
{\em superperiodicity} of the $ \psi_{n{\bf k}} $ themselves.
In the standard language of electronic-structure
calculations, if the $ \psi_{n{\bf k}} $ have ${\bf k}$'s that are restricted to 
a uniform Monkhorst-Pack mesh, they will all be periodic with a wavelength 
inversely proportional to the spacing of the mesh; this periodicity is
consequently inherited by the WFs.
For $\mathcal N$ {\bf k}-points along a
direction of the BZ, the WFs will repeat along the corresponding
direction every $\mathcal N$ cells;
therefore a mesh of {\bf k}-points needs to be dense enough to assure
that adjacent replicas of the WFs do not overlap.\\

The method described above works properly in the case of {\em isolated
groups} of bands.~\cite{note1}  On the other hand to study quantum
conductance in extended systems we often need to compute 
WFs for a subset of energy bands that are entangled or mixed
with other bands.  Most often we are interested in the
states that lie in the vicinity of the Fermi level of a conductor
in a restricted energy range.  Since the unitary transformations
$U^{({\bf k})}$ mix energy bands at each {\bf k}-point,
any arbitrary choice of states inside a prescribed
window will affect the localization properties of WFs unless 
energy gaps
effectively separate the manifold of interest from higher and lower bands.
This problem has been solved by Souza, Marzari, and Vanderbilt, introducing
an additional disentanglement procedure~\cite{ivo2} that automatically
extracts the best possible manifold of a given dimension from the states
falling in a predefined energy window. This is the generalization to
{\em entangled} or metallic cases of the maximally-localized WF
formulation.  The procedure relies on minimizing the
subspace dispersion across the Brillouin Zone, and effectively 
extracts the bands of interest from the overall band structure.  
In practice, first we select a desired number of bands in an energy window;  then
we determine the optimally-connected subspace that can be
extracted from that band structure; and finally we proceed with
a standard localization procedure inside the selected subspace, using
the same kind of spread functional $\Omega$ and of unitary matrices $U^{({\bf k})}_{mn}$.
The resulting orbitals have the same good localization properties, and
allow to apply our formalism to arbitrary systems, independently
of the insulating or metallic nature of the band manifold. It should
be stressed that the WFs obtained in the later case are not the WFs
of the occupied subspace (that would exhibit poor localization properties), but
are those of a well connected, continuous subspace that in general will
contain both occupied and unoccupied Bloch functions.

In order to calculate the conductance according to the prescriptions outlined 
in Sec.~\ref{tbmethod}, we need as an input the matrix elements of the
Hamiltonian calculated on a localized basis: in our case, it is the
minimal basis of the maximally-localized WFs. The advantages of this choice 
are twofold: firstly, besides being a minimal basis, the WFs
span {\it exactly} the Hilbert space of an insulator and, with arbitrary
accuracy, of an entangled metallic system. 
Secondly, their localization assures the choice of the system with the smallest number 
of atomic layers.
The Hamiltonian matrices ($H^{LR}_{mn}$,
$H_C$, $H_{LC}$, $H_{CR}$) can be formally obtained from the {\em on
site} ($H_{00}$) and {\em coupling} ($H_{01}$) matrices between {\em
principal layers}.  In our formalism, and assuming a BZ sampling fine
enough to eliminate the interaction with the periodic images, we 
can simply compute these matrices from the
unitary matrix $U^{({\bf k})}$ obtained in the localization
procedure.~\cite{note2} By definition of energy eigenvalues
($\widetilde{\epsilon}_{m{\bf k}}$), the Hamiltonian matrix
$\widetilde{H}_{mn}({\bf k})= \widetilde{\epsilon}_{m{\bf k}}
\delta_{m,n}$, is diagonal in the basis of the Bloch eigenstates.  We
can calculate the Hamiltonian matrix in the rotated basis, 
\begin{equation}
H^{(rot)}({\bf k})=(U^{({\bf k})})^{\dagger}\widetilde{H}({\bf
k})U^{({\bf k})}.
\label{rot}
\end{equation}
Next we Fourier transform $H^{(rot)}({\bf k})$ into a set of $N_{kp}$
Bravais lattice vectors {\bf R} within a Wigner-Seitz supercell
centered around {\bf R}=0 :
\begin{equation}
H^{(rot)}_{mn}({\bf R})=\frac{1}{N_{kp}} \sum_{\bf k} 
e^{-i{\bf k}\cdot {\bf R}}H^{(rot)}_{mn}({\bf k})=\langle w_{m{\bf 0}} 
| \widehat{H} | w_{n{\bf R}} \rangle,
\label{hrrot}
\end{equation}
where $N_{kp}$ derives from the folding of the uniform mesh of {\bf
k}-points in the BZ.  The term with {\bf R}=0 provides the {\em
on site} matrix $H_{00}=\langle w_{m{\bf 0}} | \widehat{H} | w_{n{\bf
0}}\rangle$, and the term {\bf R}=1 provides the {\em coupling} matrix
$H_{01}=\langle w_{m{\bf 0}} | \widehat{H} | w_{n{\bf 1}}\rangle$:
These are the only ingredients required for the evaluation of
the quantum conductance.

%%%%%%%%%%%%%%%%%%%%%%%%%%%%%%%%%%%%%%%%%%%%%%%%%%%%%%%%%%%%%%%%%%%%%%
%                    Bulk-like conductance                           %
%%%%%%%%%%%%%%%%%%%%%%%%%%%%%%%%%%%%%%%%%%%%%%%%%%%%%%%%%%%%%%%%%%%%%%
\section{Bulk-like conductance}\label{bulk-like}
As a first application of our method, we consider a case in which leads and
conductor (as sketched in Fig.~\ref{lcr}) are made
of the same material, and we compute
the conductance of the ideal and infinite nanostructure ({\em bulk-like
conductance}).  In this case, it is not necessary to distinguish
between conductor and lead terms and the {\em single layer} $H_{00}$
and the {\em coupling} $H_{01}$ matrices are the only necessary input.

We will focus on one-dimensional (1D) linear chains of atoms. The 
systems that have been studied most are chains of
Au,~\cite{ohnishi,yanson,nilius,todorov,portal,bahn,mehrez},
Al~\cite{sen,zheng} and C.~\cite{ravagnan,lang,larade,abdurahman} In
the following, we will discuss results for Al (Sec. \ref{alchain}) and
C (Sec. \ref{cchain}) chains.

%%%%%%%%%%%%%%%%%%%%%%%%%%%%%%%%%%%%%%%%%%%%%%%%%%%%%%%%%%%%%%%%%%%%%%
%                    Aluminum chain                                  %
%%%%%%%%%%%%%%%%%%%%%%%%%%%%%%%%%%%%%%%%%%%%%%%%%%%%%%%%%%%%%%%%%%%%%%
\subsection{Aluminum chain}\label{alchain}

An ideal and infinite Al chain is simulated using periodic boundary
conditions and a unit cell containing two aluminum atoms. A large
vacuum region ($\sim 10$~\AA) in the direction perpendicular to the
chain prevents the interaction with adjacent replicas.  A
($12\times1\times1$) grid of {\bf k}-points and 18 Ry energy cut-off
for the wavefunction expansion assure the convergence of the
electronic structure of the system.  The optimized Al-Al distance
($d=2.42$~\AA) is in very good agreement with previous DFT
investigations.~\cite{sen,zheng}

\begin{figure}
\includegraphics[clip,width=0.35\textwidth]{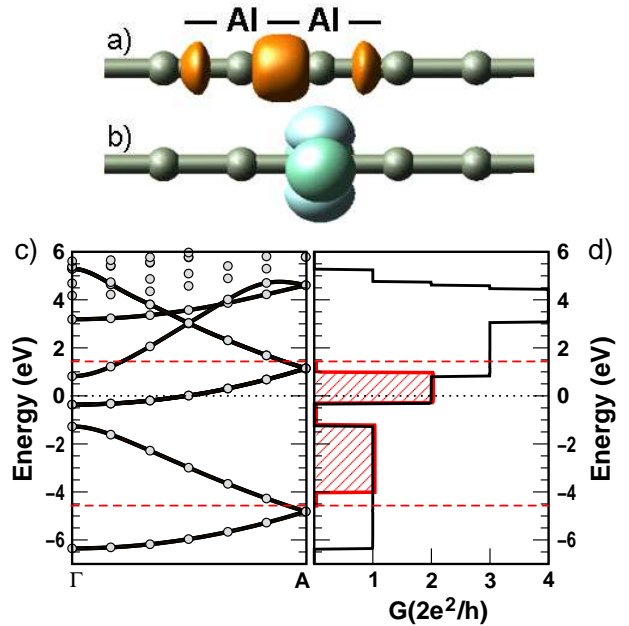}
\vspace{2mm}
\includegraphics[clip,width=0.45\textwidth]{figure2_cd.eps}
\caption{ Linear Al-chain.  Isosurface of (a) a $\sigma$ and (b) two
$\pi$ WFs. (c) Comparison between {\em original}
first-principles (grey dots) and interpolated (black lines) band
structure. 
Dotted lines represent the Fermi Level.
Note that some bands (i.e. at Fermi Level) are double degenerate.
(d) Quantum conductance plots. The solid line is relative to the whole
energy range (E $\in[-7,6]$~eV); 
the shaded area identifies the quantum conductance calculated in 
the narrow energy window (E $\in[-4.5,1.5]$~eV) bounded by the dashed lines.
}  
\label{aluminum}
\end{figure}

Following the procedure described above, we calculated the electronic
structure and quantum conductance of this system.  To construct the
WFs we selected an energy window with E $\in[-7,6]$~eV around
the Fermi level (taken as the reference zero).  
This energy window contains all
the occupied bands and the first empty states. 
We chose to extract an 8-dimensional manifold from this energy window:
After the disentanglement and localization procedure, we obtain eight WFs
which span the 8-dimensional Hilbert manifold and represent an
orthonormal minimal basis for it.  Thus, the calculation of quantum
conductance involves operations with very small ($8\times8$) matrices,
with a negligible computational effort, exactly comparable to a
tight-binding calculation (TB) with two sites and four orbital per
site. However, our results provide more
information on the electronic structure than the TB
approach.  The calculated WFs (Fig.~\ref{aluminum}a,b) are well
characterized and are consistent with the estimated chemical bonds
present in the system ($\sigma$ and $\pi$ orbitals).  The $\sigma$
states are centered in the middle of the Al-Al bond, while the $\pi$
states are localized around single atoms.

As a test of the accuracy of the WF transformation,
we have computed back the band structure of the Al-chain,
starting from the Wannier-function Hamiltonian in real space.
This Hamiltonian can now be Fourier transformed in reciprocal space
$H^{(rot)}_{mn}({\bf R})$ (Eq.~\ref{hrrot}) for any arbitrary {\bf
k}-point
\begin{equation}
H^{(rot)}_{mn}({\bf k'})= \sum_{\bf R} e^{+i{\bf k'}\cdot {\bf
R}}H^{(rot)}_{mn}({\bf R});
\label{hrotk}
\end{equation}
the resulting Hamiltonian matrices can then be diagonalized to find
energy eigenvalues.  Comparing the {\em original} PW
(grey dots) with the interpolated (black lines) band structure
(Fig.~\ref{aluminum}c) we see an excellent
agreement. This is an expected but important validation, since 
it proves that the
intermediate transformations do not affect the accuracy of the
first-principles PW calculations.  All the information on the electronic
structure of the system is transferred to the matrix elements of the
Hamiltonian expressed in the localized WFs basis.

A closer inspection to Fig.~\ref{aluminum}c shows some
unmatched electronic states at energies lower than the highest
interpolated band (in this case, in the vicinity of the 
$\Gamma$ point).  This is
the effect of the band-space minimization,~\cite{ivo2} which singles
out the best-connected manifold from an entangled group of states.  
The upper edge of the outer window 
(Fig.~\ref{aluminum}c) intersects states having 
comparable energy but different symmetries, relevant to higher parts
of the spectrum. The inclusion of these
contributions (i.e., the states around $\Gamma$ at $\sim$ 4-5 eV)
would affect significantly the localization properties of the WF
basis. The minimization of the dispersion for the extracted manifold 
(the disentanglement of the bands) is thus an essential step in the 
WFs calculation.

The disentanglement procedure can be used
to probe different energy windows;
this allows  to single out the most relevant bands,
linking conductance properties to the nature of
the chemical bonds. 
As an example, if 
we restrict the energy window to a few eV 
around the Fermi Level (dashed lines
in Figure~\ref{aluminum}c,d), we can
describe quantum conductance using the only smaller set of bands (three) included
in the narrow window. 
In this restricted range, the {\em new} conductance spectra (shaded area 
in Fig.~\ref{aluminum}d) is indistinguishable from the original one (black line).

The linear chain of Al atoms displays metallic behavior
(Fig.~\ref{aluminum}c,d), in agreement with previous DFT
calculations.~\cite{sen} Due to the reduced coordination number of
the Al atoms in the chain ($n_{chain}=2$) compared with the {\em
FCC}-bulk phase ($n_{bulk}=12$), this metallic character was not
obvious a priori.  
It is important to note that, in general, metallic systems are not
well represented in a WF framework.  So far, only transition metals 
have been the
subject of WFs studies, due to the localized character
of their {\em d}-orbitals,~\cite{sporkmann} and only recently
localized wave functions in reciprocal space have have been proposed
for simple metals (Na and Al).~\cite{parrinello} However, in the 
disentanglement procedure we are not required to restrict ourselves to the
occupied subspace, but we can mix filled and empty states, allowing
us to extract well-connected manifolds that have the same
localization properties of the manifolds for insulators and semiconductors.

To better understand the conductance properties of Al chains,
we have also calculated the eigenvectors of the transmission function
$\mathcal T$(E), generally known as {\em
eigenchannels}.\cite{jacobsen} The eigenchannels completely
characterize transmission, and at each energy, describe the single
modes of the electronic transport.  Our results show that the
eigenchannels at the Fermi energy are barely the linear combination of the
$\pi$-like WFs of Figure~\ref{aluminum}b.  The two quanta of
conductance ($2e^2/h$) at the Fermi level (Fig.~\ref{aluminum}d)
correspond to two degenerate $\pi$ states, which constitute the
channels for charge mobility. The metallic behavior is in qualitative
agreement with the geometrical properties of WFs. As mentioned
before, both $\sigma$- and $\pi$- WFs are well localized.  While
the $\sigma$ states (Fig.~\ref{aluminum}a) are centered in the
middle of the Al-Al bond, in a bonding configuration, the $\pi$
orbitals (Fig.~\ref{aluminum}b), responsible for the chain
metallicity, are centered on the single atoms.

Finally, in the absence of external leads, there is a one-to-one correspondence
between the quantum conductance spectrum (Fig. 2d) and the band
structure: at a given value of $E$, the quantum conductance
(Eq.~\ref{eq:landauer}) is a constant proportional to the number of
transmitting channels available for charge mobility, which are equal
(in a periodic system) to the number of bands at the same energy.  The
perfect agreement between band structure and quantum conductance
represents a further validation for the ability of our method to
calculate transport properties.

%%%%%%%%%%%%%%%%%%%%%%%%%%%%%%%%%%%%%%%%%%%%%%%%%%%%%%%%%%%%%%%%%%%%%%
%                    Carbon chains                                   %
%%%%%%%%%%%%%%%%%%%%%%%%%%%%%%%%%%%%%%%%%%%%%%%%%%%%%%%%%%%%%%%%%%%%%%
\subsection{Carbon chains}\label{cchain}

As a second application, we have studied two
different species of carbon chains. Nanodevices where C-chains
act as conductors bridging metal electrodes are not only ideal
prototypes for studying conduction in reduced-dimensionality systems,
but are also fundamental constituents of low-pressure carbon
assemblies, such as those found in end-capped molecules or in the
interstellar medium.
Theoretical models proposed so far have dealt with wires of
equidistant C atoms trapped between metallic leads of Au or
Al.~\cite{lang,stokbro} In those cases, the conduction properties of
the system are strongly dependent on the number (odd or even) of 
atoms in the chain.

Here, we focus on the effects of structural relaxation on the
electronic and transport properties of infinite carbon chains,
known as {\em
carbyne}. The name {\em carbyne}~\cite{fisher,kavan,gibtner} denotes an
allotrope based on a linear chain of {\em sp}-hybridized carbon atoms:
isomeric polyethynylene 
diylidene (polycumulene or cumulene) or
polyethynylene (polyyne).  The {\em cumulene} form is
characterized by an equidistant arrangement of C-atoms with double
{\em sp}-bonds $(=C=C=)_n$, while the {\em polyyne} form is a dimerized
linear chain with alternating single-triple bonds $(-C \equiv C-)_n$.
The experimental evidence for {\em carbyne} chains is controversial and its
properties not completely known.~\cite{ravagnan,kavan} We studied the
effects of the two allotropes (cumulene vs. polyyne)
to the electronic and conduction properties of {\em carbyne}.

We used four C-atoms in a
periodically repeated cell, and an ($8\times1\times1$) grid of {\bf k}-points for BZ summation.
The electronic wavefunctions are expanded in a plane-wave basis set up
to 40 Ry.  We first optimized the lattice constant of the
cumulene structure, and then, in the same unit cell, we relaxed the
carbon-carbon distances in the polyyne phase.  In the cumulene form
the C-atoms are separated by d$_{cumulene}=1.37$~\AA, while the
polyyne form dimerizes with C-atoms separated by
d$_{single}=1.51$~\AA~and d$_{triple}=1.22$~\AA, in agreement with
previous theoretical calculations.~\cite{abdurahman} Incidentally, 
polyyne  is energetically more stable than  cumulene
by 1.2 eV per unit (C-C). \noindent
\begin{figure}[!t]
\includegraphics[clip,width=0.35\textwidth]{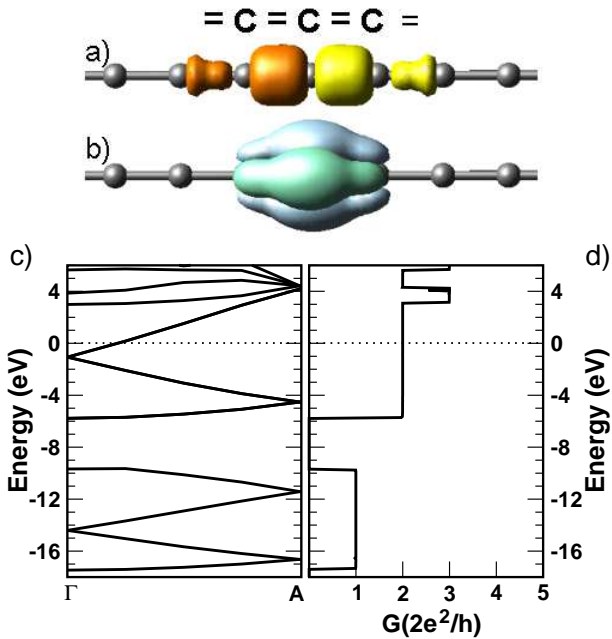}
\vspace{2mm}
\includegraphics[clip,width=0.45\textwidth]{figure3_cd.eps}
\caption{Cumulene.  Isosurface of (a) two $\sigma$ and (b) two $\pi$
WFs.  First-principles band structure (c), and  calculated quantum
conductance (d) in the selected energy window.  Dotted lines represent the
Fermi Level of the system.}
\label{cumulene}
\end{figure}

\begin{figure}[!t]
\includegraphics[clip,width=0.35\textwidth]{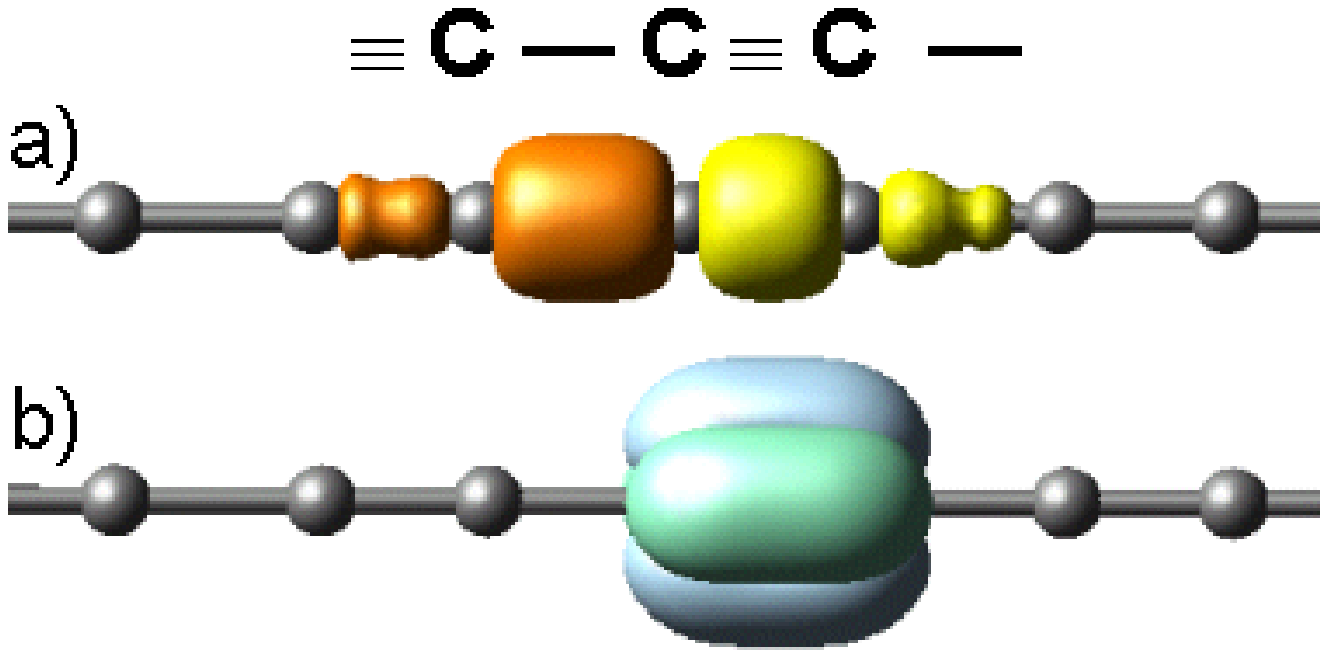}
\vspace{2mm}
\includegraphics[clip,width=0.45\textwidth]{figure4_cd.eps}
\caption{Polyyne.  Isosurface of (a) two $\sigma$ and (b) two $\pi$
WFs. 
First-principles band structure (c), and  calculated quantum
conductance (d) in the selected energy window.  Dotted lines represent the
Fermi Level of the system.}
\label{polyyne}
\end{figure}

Fig.~\ref{cumulene} and Fig.~\ref{polyyne} show our results.
Cumulene 
(Fig.~\ref{cumulene}a,b) is characterized by symmetric {\em sp}-bonds,
uniformly distributed along the chain.  $\sigma$ states are
localized in the middle of C-C bonds while $\pi$ states are
centered around single C-atoms. In polyyne (\ref{polyyne}a,b),
$\sigma$ orbitals are localized both on single $C - C$ and on triple
$C \equiv C$ bonds, with a $\sigma$ state in the middle of each
bond.  The $\pi$ orbitals are localized on the $C \equiv C$
bonds: there are two of these $\pi$ orbitals in the middle of 
each triple bond, related by a 90$^{\circ}$ rotation around the axis.

The electronic structures and quantum conductances of cumulene and
polyyne are shown in Fig.~\ref{cumulene}c,d and Fig.~\ref{polyyne}c,d
respectively.  The symmetric chain displays metallic behavior, in
agreement with previous theoretical calculations.~\cite{abdurahman} 
Polyyne is instead semiconducting: the relaxation of the
carbon-carbon distances induces a Peierls-type distortion, which
stabilizes the structure and opens energy gaps at the edges of the
Brillouin Zone. The metallicity of cumulene is an effect
related to the homogeneous distribution of the atoms, and not to the
dimensionality of the chain: polyyne, which has the same
dimensionality of cumulene, is not a metal.  As mentioned in the
previous section, the electrical (metallic or semiconducting) behavior
is tightly reflected in the geometrical properties of WFs. \noindent
The eigenchannels of both systems, near the Fermi energy, are made by
linear combination of the $\pi$-like WFs of
Figures~\ref{cumulene}-\ref{polyyne}b.  As was the case for the aluminum chain,
the $\pi$ orbitals of cumulene are located on the atoms, and the
system is metallic.  Contrary-wise, in polyyne the $\pi$ states are
centered in the middle of the triple C-bonds, and the system is
semiconducting.

%%%%%%%%%%%%%%%%%%%%%%%%%%%%%%%%%%%%%%%%%%%%%%%%%%%%%%%%%%%%%%%%%%%%%%
%                       Lead-conductor-lead                          %
%%%%%%%%%%%%%%%%%%%%%%%%%%%%%%%%%%%%%%%%%%%%%%%%%%%%%%%%%%%%%%%%%%%%%%
\section{Two terminal conductance}\label{lead-like}
\begin{figure}[!t]
\includegraphics[clip,width=0.35\textwidth]{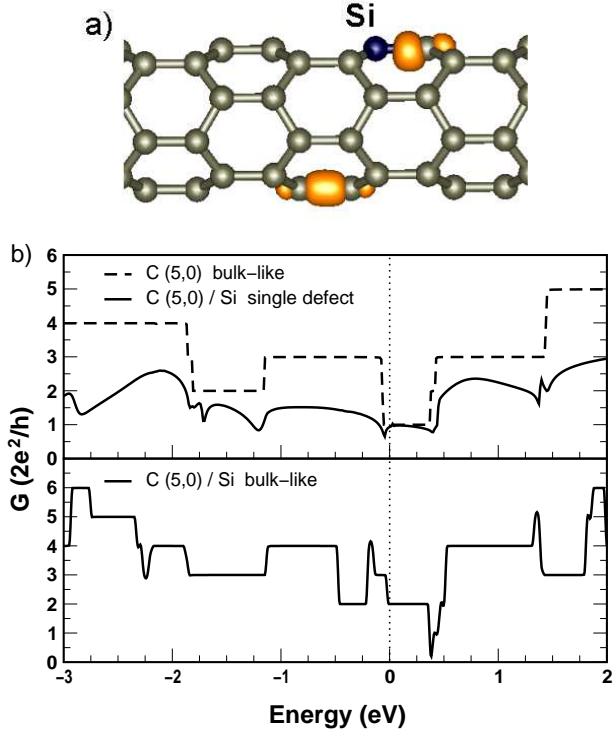}
\vspace{2mm}
\includegraphics[clip,width=0.45\textwidth]{figure5_b.eps}
\caption{ a) Isosurface of two $\sigma$ WFs in a
(5,0) carbon nanotube in presence of a substitutional Si defect (black
atom). The polarization of the $\sigma$ states is due to the effect of
the Si defect.  b) Quantum conductance plots for the (5,0) nanotube
with and without the Si defect (top panel); same nanotube with a
periodic arrangement of Si defects in a bulk-like configuration
(bottom panel).  Vertical dotted lines represent the Fermi Level of
the systems.}
\label{nano}
\end{figure}
As a prototypical example of a two terminal device we have computed
the quantum conductance for a small zigzag (5,0) carbon nanotube in the
presence of an isolated substitutional silicon defect.  Topological
and substitutional defects in carbon nanotubes have been widely
studied,~\cite{choi,chico,andriotis,kaun} and our results can be
directly compared with available theoretical data.

Using Fig.~\ref{lcr} as reference, we choose in the present case as
the {\em conductor} region a finite segment of nanotube which contains the
defect, while the {\em leads} are modeled by two semi-infinite 
nanotubes.  To compute the two terminal conductance we need, in principle,
three sets of calculations (two only if the leads are of the same
material): bulk calculations for the two infinite
leads and a supercell calculation for the conductor and the
contacts (see Fig.~\ref{lcr}).  The supercell needs to be chosen large
enough so that the influence of the conductor wavefunctions on the leads
becomes negligible at the boundaries, assuring seamless matching between the
{\em lead} WFs determined in the supercell and 
in the bulk calculations.
The matching condition can also be expressed by
saying that the {\em on-site} and {\em hopping} integrals (in the tight-binding
language) of the individual WFs have to be the identical on both
sides of the interface boundary.  
The matrices $H^{L,R}_{0,0}$ and
$H^{L,R}_{0,1}$ for the infinite leads are obtained from the bulk
calculations (in this case, the ideal (5,0)
nanotube), while the other coupling matrices are derived from the
supercell calculation. 
We stress the need to include
a sufficiently large portion of the contacts in the
supercell calculation so that, far from the interface and into the leads, the
system recovers its bulk properties. The characteristic length
determining the amount of conductor included is {\em not} the decay length
of the density matrix of the metal (long-ranged, and algebraic), but the
characteristic length of the WFs disentangled from the metal, and whose
localization properties on these well-connected manifold are comparable to
those of a semiconductor or an insulator. 
The properties of uniqueness
and localization of the WFs ensure effortless continuity across
the interface into the bulk leads. Note that a similar definition of
the geometry of the simulation cell is needed also when using other
real-space methods to compute quantum
conductance.\cite{marco2,stokbro}
Since our supercell contains both the
conductor and a portion of the leads large enough to take into account the
presence of the contacts, the Wannier transformation produces a set of WFs
covering the whole coupled region. However, since the WFs are
strongly localized, it is straightforward to distinguish those
centered on the conductor and those on the leads, 
and to see the matrix elements of the 
Hamiltonian seamless turn away from the conductor into bulk-like lead terms. 
This approach brings another 
advantage: since the conductor and the conductor-lead interfaces are simulated in
the same supercell, we have a consistent (and fully relaxed) microscopic
picture of the contacts.

The (5,0) carbon nanotube (see Fig. \ref{nano}a), has been
simulated in a ($4.25 \times 10.0 \times 10.0$)~\AA$^3$ supercell,
with 8 {\bf k}-points and a 40 Ry cutoff for the wavefunction
expansion.  Due to the highly localized nature of the 
WFs, only four atomic layers are needed to reproduce the
bulk-like behavior of the nanotube far away from the Si defect.  The
presence of the defect does not induce significant structural
relaxations, but affects the electronic and conduction properties of
the tube.  Figure~\ref{nano}a shows the isosurfaces of two
calculated WFs with $\sigma$ character.
Far away from the defect, WFs are symmetric and centered in the middle
of the C-C bond. Closer to the defect, the different electronegativity
of silicon and carbon atoms induces a polarization of the Si-C bond that
modifies the conduction properties of the system. Figure~\ref{nano}b (top panel)
displays the quantum conductance of the nanotube with respectively in the
presence of a defect (C(5,0)/Si, solid line) or in the ideal clean case
(C(5,0), dashed line).
The conductance of the ideal (5,0) nanotube shows the typical
step-like shape and a metallic behavior, which is the result of the
high curvature of such a small radius nanotube.~\cite{blase,gulseren}
Once a single Si defect is introduced, the system maintains its
metallic character but the overall spectrum changes drastically.
We observe a general reduction of the conductance along the whole
energy range, and the distortion of the step-like shape of the pure
nanotube.  The appearance of dips, corresponding to the
discontinuities in the original step function, is a characteristic
feature of nanotubes with defects,~\cite{choi,chico,kaun} where the
backscattering of electrons reduces the quantum conductance.

As a consequence of the {\em external leads} the quantum conductance
is not directly related to the  band structure of the supercell.  
We show this in Fig.~\ref{nano}b (bottom panel), where the {\em bulk} conductance of
the C(5,0)/Si system is presented.  The general trend of $\mathcal G$(E) is
different from both curves in the upper panel.  
The whole C(5,0)/Si system
is periodically repeated, and the conductance assumes the typical
step-like behavior. However, with respect to the pure nanotube,
we observe both an overall increse of the quantum conductance and 
the presence of novel features, such as the marked peak due to the silicon states,
just below the Fermi level. The periodic distribution of the Si atoms 
leads to the formation of additional channels available for the charge transport
along the tube. On the other hand, these features disappear in the case of the
single defect, where the breaking of the traslational symmetry does not 
allow the formation of delocalized orbitals.
In conclusion, while the doping with
a regular pattern of Si-atoms increases the conductance,
the scattering
dephasing of a single defect reduces the global transport
properties of the small carbon nanotube.

%%%%%%%%%%%%%%%%%%%%%%%%%%%%%%%%%%%%%%%%%%%%%%%%%%%%%%%%%%%%%%%%%%%%%%
%        Discussions and conclusions                                 %
%%%%%%%%%%%%%%%%%%%%%%%%%%%%%%%%%%%%%%%%%%%%%%%%%%%%%%%%%%%%%%%%%%%%%%
\section{Conclusions}\label{conclusions}

We have presented a novel approach to calculate quantum conductance in
extended systems in the coherent transport regime.  Our methodology
combines the accurate self-consistent minimization of the ground-state
electronic structure via first-principles calculations, the
determination of the maximally-localized WFs corresponding to manifold
of bands spanning the energy range relevant for conduction, and the
calculation of the quantum conductance using a real-space Green's
function formalism based on the Landauer approach.  This procedure
opens the way to selectively describe the quantum conductance in terms
of the relevant one-electron states that contribute directly to the
transport process.  It links the description of electronic conductance
to the intrinsic nature of the chemical bond, and gives new insights
onto the essential mechanisms that govern the electron flow at the
nanoscale. Moreover, it is computationally very efficient and can be
straightforwardly implemented as a post-processing step in any
standard electronic structure calculation~\cite{web} 
leading to a first-principles, 
highly accurate computation of
electron transport properties.

As a first illustration of the potential of this methodology, we have
studied quantum conductance in linear chains of aluminum and carbon
atoms, and in defective carbon nanotubes.  In all cases, we have
underlined the effects of the reduced coordination and of the atomic
relaxation on the transport properties and we have established a clear
relationship between the electrical characteristics and the chemical
bonds in the system.

%%%%%%%%%%%%%%%%%%%%%%%%%%%%%%%%%%%%%%%%%%%%%%%%%%%%%%%%%%%%%%%%%%%%%%
\section{Acknowledgments}

We would like to thank Dr. Paolo Giannozzi and Dr. Carlo Cavazzoni for
invaluable help and illuminating discussions.  This work was supported
in part by: MIUR (Italy) through grant FIRB-Nomade (A.\ C.); 
ONR grant N00014-01-1-1061 (N.\ M.),
the Mathematical, Information and Computational Sciences Division, 
Office of Advanced Scientific Computing Research of the U.S. 
Department of Energy under contract No. DE-AC05-00OR22725 with 
UT-Battelle and the Petroleum Research Fund of the American Chemical 
Society (M.\ B.\ N.). 
%%%%%%%%%%%%%%%%%%%%%%%%%%%%%%%%%%%%%%%%%%%%%%%%%%%%%%%%%%%%%%%%%%%%%%


\begin{thebibliography}{26}
\expandafter\ifx\csname natexlab\endcsname\relax\def\natexlab#1{#1}\fi
\expandafter\ifx\csname bibnamefont\endcsname\relax
  \def\bibnamefont#1{#1}\fi
\expandafter\ifx\csname bibfnamefont\endcsname\relax
  \def\bibfnamefont#1{#1}\fi
\expandafter\ifx\csname citenamefont\endcsname\relax
  \def\citenamefont#1{#1}\fi
\expandafter\ifx\csname url\endcsname\relax
  \def\url#1{\texttt{#1}}\fi
\expandafter\ifx\csname urlprefix\endcsname\relax\def\urlprefix{URL }\fi
\providecommand{\bibinfo}[2]{#2}
\providecommand{\eprint}[2][]{\url{#2}}
%
%
\bibitem{nygard}J. Nygard, D. H. Cobden, and P. E. Lindelof, {\em Nature} 
 {\bf408}, 342 (2000).
%
\bibitem{park}J. Park, A. N. Pasupathy, J. Goldsmith, C. Chang, Y. Yaish, J. R. 
 Petta, M. Rinkoski, J. P. Sethna, H. D. Abruna, P. L. McEuen, and D. C. Ralph,
 {\em Nature} {\bf 417}, 722 (2002).
%
\bibitem{liang}W. Liang, M. P. Shores, M. Bockrath, J. R. Long, and H. Park,
 {\em Nature} {\bf 417}, 725 (2002).
%
\bibitem{aviram}A. Aviram and M. Ratner, "Molecular Electronics: 
 Science and Technology", {\em Ann. N.Y. Acad. Sci.} {\bf 852} (1998); 
 A. Aviram, M. Ratner, and V. Mujica, "Molecular Electronics II", 
 {\em Ann. N.Y. Acad. Sci.} {\bf 960} (2002).
%
\bibitem{joachim}C. Joachim, J. K. Gimzewski, and A. Aviram, 
 {\em Nature} {\bf 408}, 541 (2000).
%
\bibitem{williams}K. A. Williams, P. T. M. Veenhuizen, B. G. de la Torre, 
 R. Eritja, and C. Dekker, {\em Nature} {\bf 420}, 761 (2002).
%
\bibitem{gudiksen}M. S. Gudiksen, L.J. Lauhon, J. Wang, D. C. Smith, and C. Lieber,
 {\em Nature} {\bf 415}, 617 (2002).
%
\bibitem{reed}M. A. Reed, C. Zhou, C. J. Muller, T. P. Burgin, and J. M. Tour, 
{\em Science} {\bf 278}, 252 (1997).
%
\bibitem{collier}C. P. Collier, E. W. Wong, M. Belohradsky, F. M. Raymo, J. F. Stoddart,
 P. J. Kuekes, R. S. Williams, and J. R. Heath, {\em Science} {\bf 285}, 391 (1999).
%
\bibitem{metzger}R. M. Metzger, {\em Acc. Chem. Res.} {\bf 32}, 950 (1999).
%
\bibitem{braun}E. Braun, Y. Eichen, U. Sivan, and G. Ben-Yoseph, {\em Nature} {\bf 391},
 775 (1998); and K. Keren, M. Krueger, R. Gilad, G. Ben-Yoseph, U. Sivan, and E. Braun,
 {\em Science} {\bf 297}, 72 (2002).
%
\bibitem{reichert}J. Reichert, R. Ochs, D. Beckmann, H. B. Weber, M. Mayor, and H.\ v. L\"ohneysen,
 {\em Phys. Rev. Lett.} {\bf 88}, 176804 (2002).
%
\bibitem{smit}R. H. M. Smit, Y. Noat, C. Untiedt, N. D. Lang, M. C. van Hemert, 
 and J. M. van Ruitenbeek, {\em Nature} {\bf 419}, 906 (2002).
%
\bibitem{lahann}J. Lahann, S. Mitragotri, T.-N. Tran, H. Kaido, J. Sundaram, I. S. Chaoi,
 S. Hoffer, G. A. Somorjai, and R. Langer, {\em Science} {\bf 299}, 371 (2003).
%
\bibitem{tans}S. J. Tans, M. H. Devoret, R. J. A. Groeneveld, and C. Dekker,
 {\em Nature} {\bf 394}, 761 (1998); and S. J. Tans and C. Dekker, {\em Nature} {\bf 404}, 834 (2000).
%
\bibitem{white}C. T. White, and T. N. Todorov, {\em Nature} {\bf 411}, 649 (2001).
%
\bibitem{ouyang}M. Ouyang, J.-L. Huang, C. L. Cheung, and C. M. Lieber,
 {\em Science} {\bf 292}, 702 (2001).
%
\bibitem{collins}P. G. Collins, M. S. Arnold, and P. Avouris, {\em Science} {\bf 292}, 706 (2001).
%
\bibitem{agrait}N. Agra\"it, A. Levy Yeyati, and J. M. van Ruitenbeek, {\em Phys. Rep.} {\bf 377}, 
 81 (2003).
%
\bibitem{ohnishi}H. Ohnishi, Y. Kondo, and K. Takayanagi, {\em Nature} {\bf 395}, 780 (1998).

%
\bibitem{yanson}A. I. Yanson, G. Rubio Bollinger, H. E. van den Brom, N. Agra\"it, and J. M. van Ruitenbeek,
 {\em Nature} {\bf 395}, 783 (1998).
%
\bibitem{nilius}N. Nilius, T. M. Wallis, and W. Ho, {\em Science} {\bf 297}, 1853 (2002); and 
 T. M. Wallis, N. Nilius, and W. Ho, {\em Phys. Rev. Lett.} {\bf 89}, 236802 (2002).
%
\bibitem{landauer}R. Landauer, {\em Philos. Mag.} {\bf 21} 863 (1970).
%
\bibitem{datta}S.\ Datta, {\em Electronic transport in mesoscopic systems} (Cambridge Univ. Press 1995).
%
\bibitem{marco}M.\ Buongiorno Nardelli, {\em Phys. Rev. B} {\bf 60}, 7828 (1999).
%
\bibitem{todorov}T. N. Todorov, J. Hoekstra, and A. P. Sutton, {\em Phys. Rev. Lett.} {\bf 86}, 3606 (2001).
%
\bibitem{kristic}P. S. Krstic, X.-G. Zhang, and W. H. Butler,  {\em Phys. Rev. B} {\bf 66}, 205319 (2002).
%
\bibitem{rego}L. G. C. Rego, A. R. Rocha, V. Rodrigues, and D. Ugarte, {\em Phys. Rev. B} {\bf 67}, 045412 (2003).
%
\bibitem{lang1}N. D. Lang, {\em Phys. Rev. B} {\bf 52}, 5335 (1995); and
 M. Di Ventra, S. T. Pantelides, and N. D. Lang, {\em Phys. Rev. Lett.} {\bf 84}, 979 (2000).
%
\bibitem{choi}H. J. Choi and J. Ihm, {\em Phys. Rev. B} {\bf 59}, 2267 (1999); and
  H.J. Choi and J. Ihm and S.G. Louie and M.L. Cohen, {\em Phys. Rev. Lett.} {\bf 84}, 2917 (2000).
%
\bibitem{landman}U. Landman, {\em Proceedings of the Workshop on R\&D Status and Trends in Nanoparticles, 
 Nanostructured Materials, and Nanodevices in the United States}, NSF, p. 192 (1997).
%
\bibitem{landman1}U. Landman,  R. N. Barnett,  A. G. Scherbakov, and P. Avouris, 
 {\em Phys. Rev. Lett.} {\bf 85} 1958 (2000).
%
\bibitem{yoon}Y.-G. Yoon, M. S. C. Mazzoni, H. J. Choi, J. Ihm, and S. G. Louie, 
 {\em Phys. Rev. Lett.} {\bf 86}, 688 (2001).
%
\bibitem{guo}J. Taylor, H. Guo, and J. Wang, {\em Phys. Rev.  B} {\bf  63}, 245407 (2001).
%
\bibitem{yaliraki}S. N. Yaliraki, A. E. Roitberg, C. Gonzalez, V.  Mujica, and M. A. Ratner, 
 {\em J. Chem. Phys.} {\bf 111}, 6997 (1999).
%
\bibitem{stokbro} M. Brandyge, J.-L. Mozos, P. Ordejon, J. Taylor, and K. Stokbro, 
 {\em Phys. Rev. B} {\bf 65}, 165401 (2002).
%
\bibitem{palacios} J. J. Palacios, A. J. P\'erez-Jemenez, E. Luis, E. SanFabi\'an, and J. A. Verg\'es, 
 {\em Phys. Rev. B} {\bf 66} 035322 (2002).
%
\bibitem{datta1} S.\ Datta, {\em Superlatticess and Microstructures} {\bf 28}, 253 (2000); 
 and Y. Xue, S.\ Datta, and  M. A. Ratner, {\em J. Chem. Phys.} {\bf 115}, 4292 (2001).
%
\bibitem{arrigo}A. Calzolari, R. Di Felice, E. Molinari, and A. Garbesi, {\em Appl. Phys. Lett.} {\bf 80}, 3331 (2002).
%
\bibitem{parrinello1}F. L. Gervasio, P. Carloni, and M. Parrinello, {\em Phys. Rev. Lett.} 
 {\bf 89}, 108102 (2002). 
%
\bibitem{rosa}R. Di Felice, A. Selloni, and E. Molinari,  {\em J. Phys. Chem B} {\bf 107}, 1151 (2003). 
%
\bibitem{marco1} M. Buongiorno Nardelli and J. Bernholc, {\em Phys. Rev. B} {\bf 60}, R16338 (1999).
%
\bibitem{marco2}M. Buongiorno Nardelli, J.-L. Fattebert, and J. Bernholc, {\em Phys. Rev.  B}
 {\bf  64}, 245423 (2001)
%
\bibitem{nicola}N. Marzari and D. Vanderbilt, {\em Phys. Rev. B} {\bf 56}, 12847 (1997).
%
\bibitem{lee}D. Lee and J. Jannopoulos, {\em Phys. Rev. B}  {\bf 23}, 4988 (1981).
%
\bibitem{garcia}F. Garcia-Moliner and V. R. Velasco, {\em Theory of Single and Multiple Interfaces}, 
 (World Scientific, Singapore, 1992).
%
\bibitem{note_metal} For the case of a metal we need to
disentangle the states inside a given energy window from the rest of the
Hilbert space, using the method described in Section~\ref{wannier_theo}.
%
\bibitem{dreizler} R. M. Dreizler and E. K. U. Gross, {\em Density Functional Theory. An approach
  to the quantum many-body problem}, (Springer-Verlag, Berlin 1990).
%
\bibitem{pederw}  D. M. Ceperley and B. J. Alder, {\em Phys. Rev. Lett.} {\bf 45}, 566 (1980);
  J. P. Pederw and A. Zunger, {\em Phys. Rev. B} {\bf 23}, 5048 (1981).
%
\bibitem{tm} N. Troullier and J. L. Martins, {\em Phys. Rev. B} {\bf 43}, 1993 (1991).
%
\bibitem{kb} L. Kleinman and D. M. Bylander, {\em Phys. Rev. Lett.} {\bf 48}, 1425 (1982).
%
\bibitem{mp} H. J. Monkhorst and J. D. Pack, {\em Phys. Rev. B} {\bf 13}, 5188 (1976).
%
\bibitem{jones}R. O. Jones and O. Gunnarsson, {\em Rev. Mod. Phys.} {\bf 61}, 689 (1989). 
%
\bibitem{wimmer} E. Wimmer, {\em Density Functional Approaches for Molecular and Materials Design}, 
 ACS Symposium Series {\bf 629}, p. 423, edited by B. B. Laird, R. B. Ross,
 and T. Ziegler, American Chemical Society, Washington DC (1996).
%
\bibitem{LeeMBNMarzari} Y. Lee, M. Buongiorno Nardelli and N. Marzari, to be published.
%
\bibitem{boys}S. F. Boys, {\em Rev. Mod. Phys.}, {\bf 32}, 300 (1960).
%
\bibitem{Silvestrelli}
P. L. Silvestrelli, N. Marzari, D. Vanderbilt, and M. Parrinello,
 {\em Sol. Stat. Comm.}, {\bf 107}, 7 (1998).
%
\bibitem{kohn} W. Kohn, {\em Phys. Rev.} {\bf 115}, 809 (1959); and
 M. R. Geller and W. Kohn, {\em Phys. Rev. B} {\bf 48}, 14085 (1993).
%
\bibitem{note_ref}The minimization of the spread functional
$\Omega$ is obtained via a steepest descent scheme, in the reciprocal space. This procedure,
computationally inexpensive, requires the updating  only of the overlap matrices M$^{({\bf k},{\bf b})}_{mn}$,
which appear in the formulation of the functional $\Omega$ (M$^{({\bf k},{\bf b})}$) 
and of the matrices U$^{{\bf k}}$(M$^{({\bf k},{\bf b})}$). For the complete description
of the method see Ref. [44]. 
%
\bibitem{note1}A group of bands are said to form
a {\em isolated composite group} if they are inter-connected by degeneracy, but are {\em
isolated} from all the other bands.
%
\bibitem{ivo2}I. Souza, N. Marzari, and D. Vanderbilt, {\em Phys. Rev. B} {\bf 65}, 035109 (2002).
%
\bibitem{note2} $U_{mn}^{({\bf k})}$  is the result of the minimization 
procedure of Eq. \ref{minimo}; we dropped the suffix $ML$ to simplify 
the notation.
%
\bibitem{portal}D. S\'anchez-Portal, E. Artacho, J. Junquera, P. Ordej\'on,  A. Garc\'ia, and J. M. Soler,
 {\em Phys. Rev. Lett.} {\bf 83}, 3884 (1999); and 
D. S\'anchez-Portal, E. Artacho, J. Junquera, A. Garc\'ia, 
 and J. M. Soler, {\em Surf. Sci.} {\bf 482}, 1261 (2001).
%
\bibitem{bahn}S. R. Bahn and K. W. Jacobsen, {\em Phys. Rev. Lett.} {\bf 87}, 26101 (2001); and 
 S. R. Bahn, N. Lopez, J. K. Norskov, and K. W. Jacobsen, {\em Phys. Rev. B} {\bf 66}, R081405 (2002).
%
\bibitem{mehrez} H. Mehrez, A. Wlasenko, B. Larade, J. Taylor, P. Gr\"utter, and H. Guo,
 {\em Phys. Rev. B} {\bf 65}, 195419 (2002).
%
\bibitem{sen}P. Sen, S. Ciraci, A. Buldum, and P. Batra,  {\em Phys. Rev. B} {\bf 64}, 195420 (2001).
%
\bibitem{zheng}J.-C. Zheng, H.-Q. Wang, A. T. S. Wee, and C. H. A. Huan, {\em Int. J. Nanosci.} {\bf 1}, 159 (2002).
%
\bibitem{ravagnan}L. Ravagnan, F. Siviero, C. Lenardi, P. Piseri, E. Barborini, P. Milani, 
 C. S. Casari, A. Li Bassi, and C. E. Bottani, {\em Phys. Rev. Lett.} {\bf 89}, 285506 (2002).
%
\bibitem{lang}N. D. Lang and Ph. Avouris, {\em Phys. Rev. Lett.} {\bf 81}, 3515 (1998).
%
\bibitem{larade}B. Larade, J. Taylor, H. Mehrez and H. Guo,  {\em Phys. Rev. B} {\bf 64},
 075420 (2001). 
%
\bibitem{abdurahman}A. Abdurahman, A. Shukla, and M. Dolg, {\em Phys. Rev. B} {\bf 65}, 115106 (2002).
%
\bibitem{sporkmann}B. Sporkmann and H. Bross, {\em Phys. Rev. B} {\bf 49}, 10869 (1994).
%
\bibitem{parrinello}M. Iannuzzi and M. Parrinello, {\em Phys. Rev. B} {\bf 66}, 155209 (2002).
%
\bibitem{jacobsen} M. Brandbyge, M. Sorensen, and K. Jacobsen,
Phys. Rev. B {\bf 56}, 14956 (1997).
%
\bibitem{fisher}E. O. Fischer, {\em On the road to carbene and carbyne complexes}, Nobel Lecture (1973).
%
\bibitem{kavan}L. Kavan, {\em Carbon} {\bf 36}, 801 (1998), and references therein.
%
\bibitem{gibtner}T. Gibtner, F. Hampel, J. P. Gisselbrecht, and A. Hirsh, {Chem. Eur. J.} {\bf 8}, 408 (2002).
%
\bibitem{chico}L. Chico, L. X. Benedict, S. G. Louie, and M. Cohen, {\bf 54}, 2600 (1996).
%
\bibitem{andriotis}A. N. Andriotis, M. Menon, and D. Srivastava, {\em J. Chem. Phys.} {\bf 117}, 2836 (2002).
%
\bibitem{kaun}C.-C. Kaun, B. Larade, H. Mehrez, J. Taylor, and H. Guo, {\em Phys. Rev. B} {\bf 65}, 205416 (2002).
%
\bibitem{blase}X. Blase, L. X. Benedict, E. L. Shirley, and S. G. Louie, {\em Phys. Rev. Lett.} {\bf 72}, 1878 (1994).
%
\bibitem{gulseren}O. G\"ulseren, T. Yildirim, and S. Ciraci, {\em Phys. Rev. B} {\bf 65}, 153405 (2002).
%
\bibitem{web} The wannier code will soon be publicly released at
http://www.wannier.org . 
\end{thebibliography}
\end{document}